\documentclass[prl,twocolumn,superscriptaddress,amsmath,amssymb,floats]{revtex4}
\usepackage{tipa}

\usepackage{graphicx}

\begin{document}

\title{Superconducting coherence peak in the electronic excitations of a single layer cuprate superconductor Bi$_{2}$Sr$_{1.6}$La$_{0.4}$CuO$_{6+\delta}$}

\author{J.~Wei}
\author{Y.~Zhang}
\author{H.~W.~Ou}
\author{B.~P.~Xie}
\author{D.~W.~Shen}
\author{J.~F.~Zhao}
\author{L.~X.~Yang}
\affiliation{Department of Physics, Surface Physics Laboratory
(National Key Laboratory) and Advanced Materials Laboratory, Fudan
University, Shanghai 200433, P. R. China}

\author{M.~Arita}
\author{K.~Shimada}
\author{H.~Namatame}
\author{M.~Taniguchi}
\affiliation{Hiroshima Synchrotron Radiation Center and Graduate
School of Science, Hiroshima University, Hiroshima 739-8526, Japan}

\author{Y.~Yoshida}
\author{H.~Eisaki}
\affiliation{National Institute of Advanced Industrial Science and
Technology (AIST), Tsukuba, Ibaraki, 3058568, Japan}

\author{D.~L.~Feng}
\email{dlfeng@fudan.edu.cn} \affiliation{Department of Physics,
Surface Physics Laboratory (National Key Laboratory) and Advanced
Materials Laboratory, Fudan University, Shanghai 200433, P. R.
China}

\date{\today}

\begin{abstract}
Angle resolved photoemission spectroscopy study is reported on a
high quality optimally doped
Bi$_{2}$Sr$_{1.6}$La$_{0.4}$CuO$_{6+\delta}$ high-$T_c$
superconductor. In the antinodal region with maximal $d$-wave gap,
the symbolic superconducting coherence peak, which has been widely
observed in multi-CuO$_{2}$-layer cuprate superconductors, is
unambiguously observed in a single layer system. The associated
peak-dip separation is just about 19 meV, which is much smaller than
its counterparts in multi-layered compounds, but correlates with the
energy scales of spin excitations in single layer cuprates.
\end{abstract}

\pacs{74.72.Hs, 71.18.+y, 79.60.Bm}

 \maketitle


In search of the mechanism of high temperature superconductivity,
one central issue is whether there are certain bosons that play the
critical mediating role of phonons in conventional BCS
superconductor. Specifically, if there were such bosons, what would
they be? So far, signatures of electron-boson interactions have been
identified in single particle excitations measured by angle resolved
photoemission spectroscopy (ARPES). For example, in the so-called
nodal region, where the $d$-wave superconducting gap diminishes, a
characteristic kink\cite{ZXReview,CukReview} was discovered in the
dispersion of various cuprate superconductors. However, the nature
of the corresponding boson, particularly whether it is due to
lattice or spin excitations, was intensively debated.

For multi-CuO$_2$-layer cuprate superconductors, such as
Bi$_2$Sr$_2$CaCu$_2$O$_{8+\delta}$ (Bi2212),
Bi$_2$Sr$_2$Ca$_2$Cu$_3$O$_{10+\delta}$ (Bi2223) and
YBa$_2$Cu$_3$O$_{7-\delta}$ (YBCO)
\cite{Dessau1991,DlScience,DHLu2001,DL2002,Matsui2003},  a sharp
peak emerges out of the normal state broad spectrum when temperature
is lowered below the superconducting transition temperature ($T_c$),
forming a so-called ``peak-dip-hump''(PDH) structure (see Fig. 4c
for illustration) in the antinodal region, where $d$-wave
superconducting gap is at its maximum. This sharp peak, also termed
as superconducting coherence peak (SCP), contains rich information
about superconductivity. Its position reflects the maximal pairing
strength, while its intensity grows with decreasing temperature,
illustrating the gain of coherence through the development of
superconducting condensate or superfluid
\cite{DlScience,RHHe2004,Ding2001}. The PDH structure has been
regarded as an evidence for interactions with a bosonic mode
\cite{Campuzano1999,Norman1998,ZXShen1997,Devereaux2004,TCuk2004,Abanov1999,Eschrig2002,Zasadzinski2001,Lee2006}.
However, since this mode energy ($\sim$ 35 meV) coincides with the
energy of certain oxygen phonon and spin excitations near
($\pi,\pi$), scenarios based on both kinds of bosons have been put
forward.

\begin{figure*}[t]
\includegraphics[width=15cm]{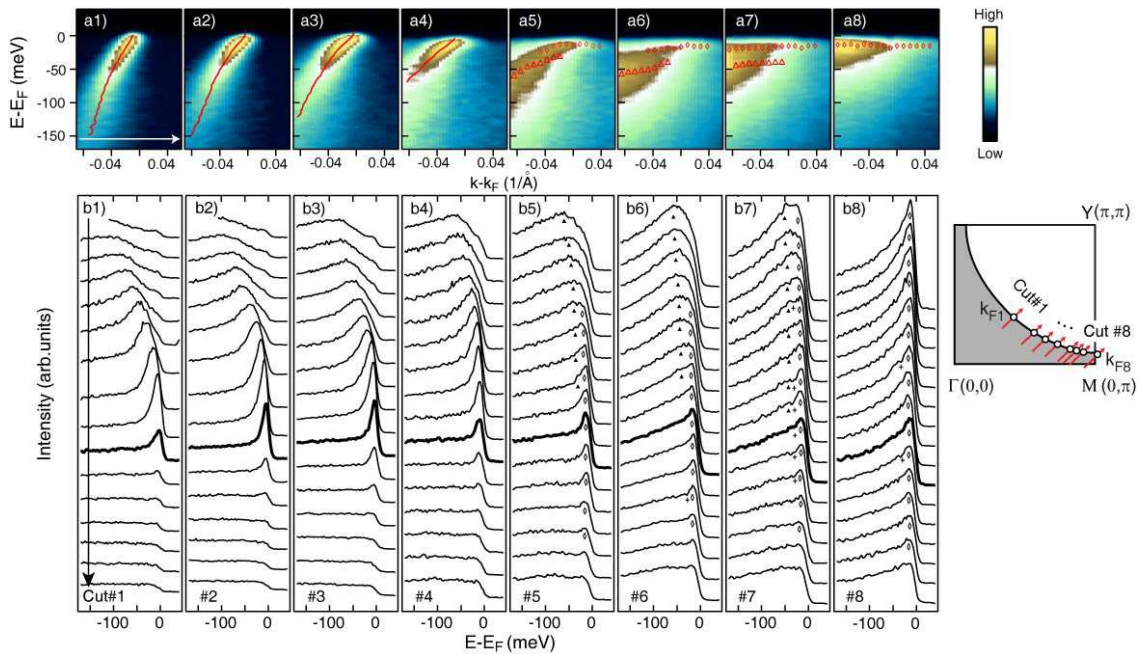}
\caption{(Color online) Photoemission data of La-Bi2201. (a1-a8)
photoemission intensity map along cut \#1-\#8 across the Fermi
surface, as illustrated by arrows on the right. The data were taken
at 20K in the superconducting state. Thick lines in a1-a4 are the
dispersions fitted from momentum distribute curves. (b1-b8) display
the corresponding spectra in panel a1-a8. The spectra at various
Fermi crossings ($k_{Fi}$,${i=1,2,\ldots,8}$, marked by black dots)
are in thick curves. The ``peak'', ``dip'' and ``hump'' structures
are represented by diamonds, crosses and triangles respectively.}
\label{f1}
\end{figure*}

In spite of its importance, an unambiguous observation of SCP or
two-component spectrum in the antinodal region of single layer
compounds is so far lacking
\cite{Kondo2007,Terashima2007,DL2002,Plate2005,LeeArx}. In this
letter, we report the discovery of the antinodal SCP in an optimally
doped single layer Bi$_{2}$Sr$_{2}$CuO$_{6+\delta}$ (Bi2201). We
found the peak-dip distance to be about 19 meV, suggesting the
possibility of a bosonic mode in Bi2201 with a much lower energy
than that in Bi2212. This low energy mode seems to correlate with
the superconducting gap and the energy scale of spin excitations
observed in various single layer compounds.

Bi$_2$Sr$_{1.6}$La$_{0.4}$CuO$_{6+\delta}$ (La-Bi2201) single
crystals were grown with the floating zone technique. Its superior
quality is manifested through the very high $T_c$ (34K) in its
class, and a remarkable zero residual resistivity
\cite{Fujita2005,Eisaki2004}, which is critical for revealing the
intrinsic properties of high-$T_c$ superconductors. The sample is
optimally doped, as verified by the linear temperature dependence of
resistivity, and detailed annealing studies. The ARPES experiments
were conducted with 22.5\,eV photons and a Scienta-R4000 electron
analyzer at beamline 9 of Hiroshima synchrotron radiation center.
The energy resolution was 7\,meV. The samples were cleaved in
ultra-high vacuum ($\sim 5\times 10^{-11}\,mbar$), and measured
within 10 hours. The experiments on Bi2212 and Bi2223 were conducted
with 22.7\,eV photons and a Scienta-200 electron analyzer at
beamline 5-4 of Stanford synchrotron radiation laboratory (SSRL).
The energy resolutions was 10\,meV. Angular resolution was
 0.3\textordmasculine for all experiments.


Photoemission data in the superconducting state of La-Bi2201 are
shown in Fig. 1 along eight cuts across the Fermi surface. The data
were taken in the $\Gamma$-Y quadrant of the Brillouin zone, so that
superstructure band is well separated from the main band. Sharp
quasiparticle shows clear dispersion in the nodal region [Fig.
1(a1,b1)]. The departure from a linear dispersion, or a kink, in the
energy range of 40$\sim$70 meV could be observed in the first three
cuts [Fig. 1(a1-a3)]. The kink diminishes further away from the
nodal region [Fig. 1(a4)], while starting from cut \#5 in the
antinodal region, a feature near the Fermi energy ($E_F$) gradually
becomes quite prominent as indicated by the diamonds in Fig.
1(a5-a8, b5-b8). Along the cuts \#5-\#7, there is also a broad
feature at higher binding energy as labeled by the triangles. Though
weak, the appearance of such two-component structure over a broad
momentum area clearly proves its robustness against the experimental
uncertainties. Eventually, a weak PDH feature is observed in some
spectra. In the vicinity of ($\pi$,0) [Fig. 1(a8,b8)], the overall
lineshape becomes sharper and closer to the $E_F$, so the two
components become indistinguishable. We emphasize the sharp peak in
the antinodal region is so delicate that it disappears further away
from the Fermi surface, where the scattering becomes stronger.
Therefore, we attribute the discovery of the two-component lineshape
to the minimized disorder effects in La-Bi2201
\cite{Fujita2005,Eisaki2004}. We also note that the evolution from
the nodal kink to the antinodal PDH  in Bi2201 is quite different
from that in Bi2212. For Bi2212, the kink becomes increasingly
pronounced away from the node, and gradually evolves into the PDH in
the antinodal region \cite{CukReview,TCuk2004}. This highlights the
subtle differences between single layer and multi-layer cuprates.


The two-component lineshape in the antinodal region alludes to the
existence of SCP in La-Bi2201. To further evidence this, the
temperature dependences along the cut \#1 and \#6 are compared in
Fig. 2. While the nodal kink structure is temperature independent
[Fig. 2(a1-a3)], a sharp peak along cut \#6 gradually emerges out of
the normal state broad spectra with decreasing temperature [Fig.
2(b1-b3)], which is more discernible in the energy distrubution
curves (EDCs) shown in Fig. 2(c1-c3). As further illustrated in Fig.
3(a-b) for two momenta in the antinodal region, the sharp peak
appears only below $T_c$, and saturates at low temperatures, just
like the temperature dependence of a SCP\cite{DlScience}. We note
that the formation of the dip feature is particularly distinct here
as well. Moreover, to show this sharp peak is not simply due to
thermal effect, the normal state spectrum at $k_{F7}$ is divided by
the resolution convoluted Fermi function of 35K and then multiplied
by that of 10K, before it is compared with the corresponding 10K
spectrum in Fig. 3c. The difference is very clear, and sharp peak
thus cannot be reproduced from thermal sharpening. For comparison,
following the same procedure, the nodal spectra taken at both below
and above $T_c$ overlap with each other (Fig. 3d). Therefore, both
the temperature and momentum dependence evidence that the sharp peak
in the antinodal region is the SCP as observed in multi-layered
cuprates before. It unambiguously rules out scenarios that suggest
PDH being caused by bilayer band splitting.

\begin{figure}[t]
\includegraphics[width=8.5cm]{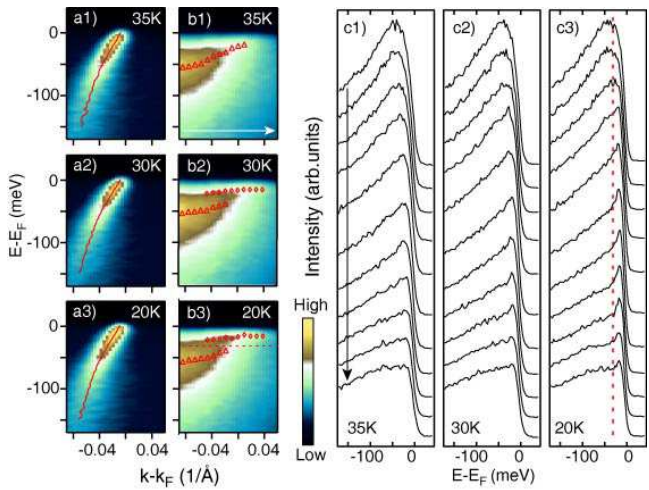}
\caption{(Color online) Temperature dependence of La-Bi2201
($T_c$=34K) photoemission data along (a1-a3) the nodal direction,
and (b1-b3) along cut \#6. (c1-c3) The EDCs of data in panel b1-b3
respectively. The red dashed line indicates the break between the
peaks (diamonds) and humps (triangles).} \label{f2}
\end{figure}

The properties of the SCP provide important information about
superconductivity. Fig. 4(a) shows the associated PDH structures for
three momenta near $(\pi,0)$. Through a phenomenological fitting
\cite{DlScience}, one could precisely determine the peak position
and the separation between the peak and dip, which is about 19 meV
at all three momenta. Similarly, the peak dip structure in Bi2212
($T_c=90K$) also possesses a constant separation of 33 meV at these
momenta (Fig. 4b), and $(\pi,0)$ (Fig. 4c). On the other hand, the
peak-dip separations are about 34 meV for Bi2223 ($T_c=108K$) and
the recently synthesized Bi2212 ($T_c=96K$). If one would follow the
same analyses as in many previous ARPES studies that the peak-dip
separation corresponds to the energy of a bosonic mode
\cite{Campuzano1999,Norman1998,ZXShen1997,Devereaux2004,TCuk2004,Abanov1999,Eschrig2002,Zasadzinski2001},
our Bi2201 data would suggest a bosonic mode of about 19 meV that
interacts with the electrons in the antinodal regions. Such a low
energy mode could not be one of the oxygen-related phonons, which
are all above 35 meV as shown by Raman scattering
experiments\cite{RLiu1992}. In the phonon picture, one then needs
different phonons to explain the PDH in Bi2212 and Bi2201. The
oxygen $B_{1g}$-phonon, which was argued to be responsible for the
PDH in Bi2212\cite{Devereaux2004,TCuk2004}, is forbidden by symmetry
in Bi2201. Phonons of heavy ions are needed to explain the low
energy feature in Bi2201\cite{RLiu1992}. On the other hand, the
peak-dip separations have an intriguing correlation with the energy
scale of spin excitations near (${\pi}$,${\pi}$) measured by
inelastic neutron scattering on YBCO \cite{BokArxiv}, Bi2212
\cite{Fong1999,HHe2001}, Pr$_{0.88}$LaCe$_{0.12}$CuO$_{4-\delta}$
(PLCCO) \cite{Wilson2007}, and La$_{2-x}$Sr$_{x}$CuO$_{4}$ (LSCO)
\cite{LSCO2007} as a function of  $T_c$ (Fig. 4d). Although
inelastic neutron scattering measurement of Bi2201 is currently
lacking, the Bi2201 peak-dip separation falls in the dashed line
representing the energy scale of the spin excitations. Moreover, the
temperature dependence of the dip structure also resembles that of
the neutron resonance mode \cite{RHHe2004,Zasadzinski2001,HHe2001}
which emerges only below $T_c$ and develops into a sharp collective
spin excitation at very low temperatures. The weaker dip structure
in Bi2201 is consistent with the weaker intensity of the spin
excitations observed in low-$T_c$ systems
\cite{Wilson2007,LSCO2007}. Furthermore, as shown in Fig. 4e, the
Bi2201 SCP position, i.e., the superconducting gap amplitude,
together with those of other optimally doped cuprates, correlates
with $T_c$, and the energy scale of the spin excitations. This
resembles the correspondence among the Debye frequency of phonon,
the superconducting gap, and $T_c$ in the BCS theory.

\begin{figure}[t]
\includegraphics[width=7cm]{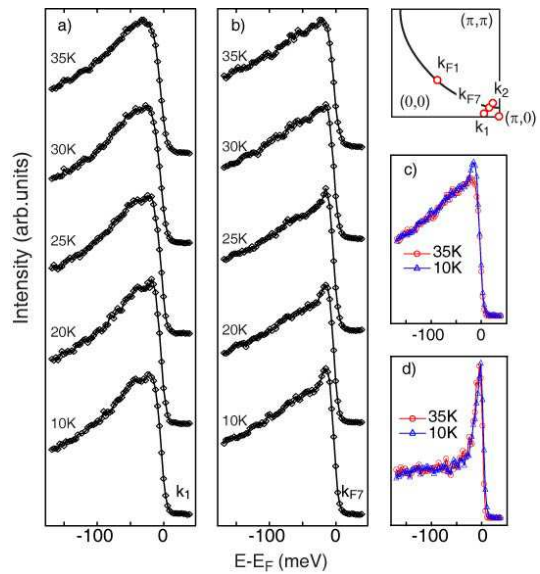}
\caption{(Color online) Detailed temperature dependence of
photoemission data at (a) $k_1$ and (b) $k_{F7}$ respectively, in
the vicinity of ($\pi,0$) as indicated in the inset. (c) and (d)
compare the spectra below and above $T_{c}$ at $k_{F7}$ and $k_{F1}$
respectively, after removing the temperature broadening effects (see
text for details). } \label{f3}
\end{figure}

Recent analysis of the Bi2212 scanning tunneling microscopy (STM)
data \cite{Lee2006} suggested that the mode energy could be the
separation between the peak and the higher binding energy edge of
the dip in the measured density of state, due to the complication of
the scattering matrix element and detailed energy and momentum
distribution of the states. In this regard,  further theoretical
analysis is necessary to conclude if the peak-dip separation in
ARPES antinodal spectrum exactly corresponds to the bosonic mode
energy. Literally following the STM analysis on antinodal spectra,
we obtained a slightly larger mode energy of about 23\,meV. From a
different perspective, the two-component lineshapes, particularly
the breaks in dispersions between the peak and hump features [e.g.,
the red dashed lines in Fig. 2(b3) and (c3)], are also possible
signs of interactions between electrons and bosons. The distance
between the peak position at the Fermi momentum and the top of the
hump feature gives an upper limit of 25\,meV for the mode. These all
robustly and self-consistently indicate a low energy mode active in
the antinodal region.

\begin{figure}[t]
\includegraphics[width=8.5cm]{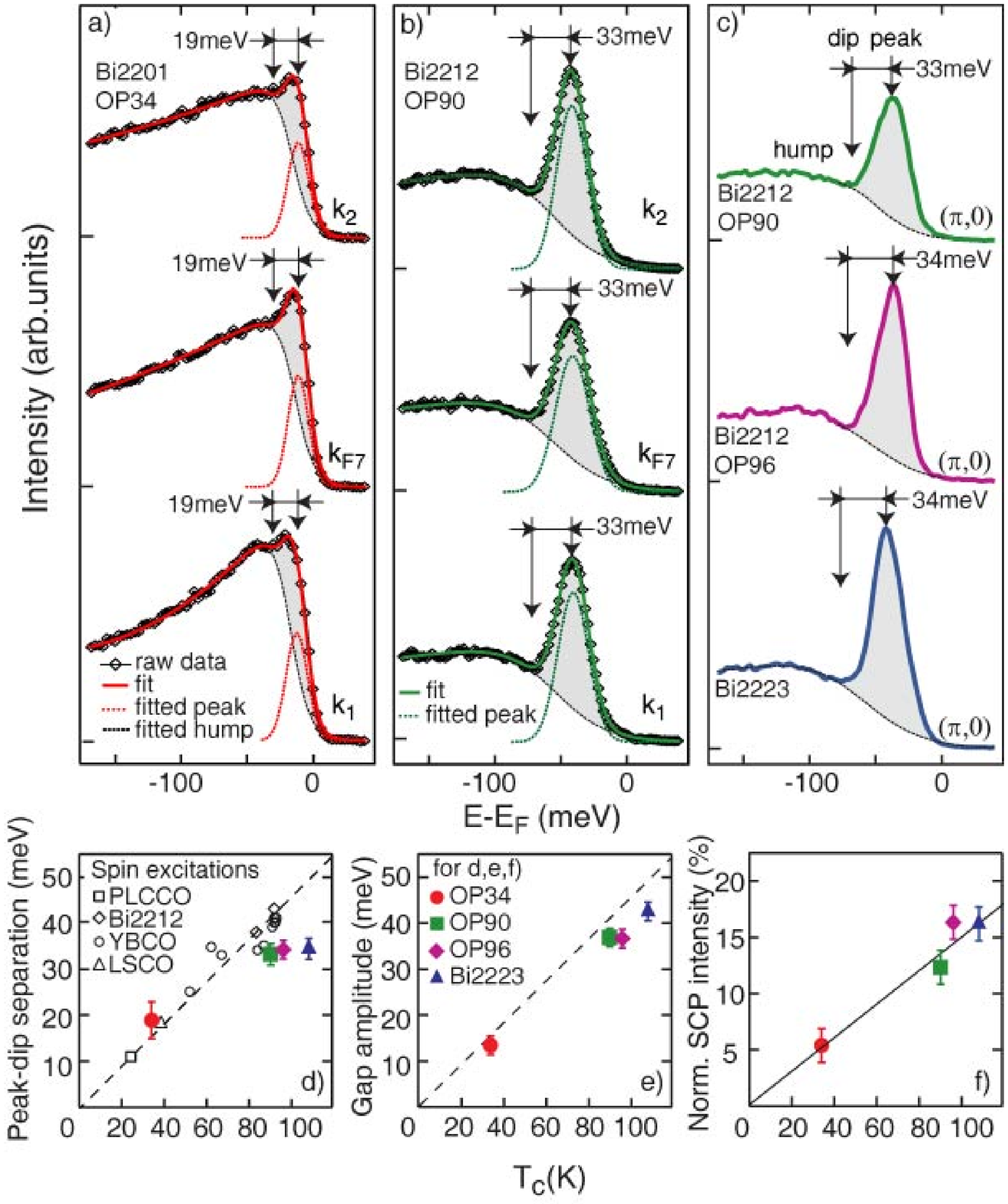}
\caption{(Color online) SCP near ($\pi$,0) for (a) optimally doped
single layer Bi$_2$Sr$_{1.6}$La$_{0.4}$CuO$_{6+ \delta}$ ($T_c$=34K,
OP34) and (b) bilayer Bi$_{2.1}$Sr$_{1.9}$CaCu$_{2}$O$_{8+\delta}$
($T_c$=90K, OP90) at three momenta ($k_{1}$ , $k_{2}$ , and $k_{F7}$
are illustrated in the inset of Fig. 3). (c) Spectra at ($\pi$,0)
for OP90, Bi$_2$Sr$_2$Ca$_{0.92}$Y$_{0.08}$Cu$_2$O$_{8+\delta}$
($T_c$=96K, OP96) and Bi2223 ($T_c$=108K). Peak and dip positions
are illustrated by the arrows. (d) The peak-dip separation vs.
$T_{c}$ for single, bi- and tri-layer bismuth cuprates near optimal
doping. The energies of spin exciations near ($\pi,\pi$) for YBCO
\cite{BokArxiv}, Bi2212 \cite{Fong1999,HHe2001}, PLCCO
\cite{Wilson2007} and LSCO \cite{LSCO2007} are plotted for
comparison. (e) The superconducting gap amplitude, and (f) the
intensity of SCP (shaded region in a,c) normalized by the total
spectral weight over [$E_F$-0.5eV, $E_F$+0.1eV] as a function of
$T_c$. The SCP is determined from a phenomenological fitting (dashed
curves) \cite{DlScience,DL2002}. The dashed lines in (d-e) are a fit
to the spin excitation energy. The solid line in (f) is a guide to
the eye.}
\end{figure}


Besides the interpretation of electron-boson interactions for PDH,
our data seem also fit in an alternative scenario, where the SCP was
considered as the emergence of a coherent quasiparticle out of the
broad incoherent background upon superfluid
formation\cite{DlScience}. The SCP intensity was found to correlate
almost linearly with the superfluid density for Bi2212, as a
signature for High-$T_c$ superconductor being a doped Mott
insulator\cite{DlScience,RHHe2004,Uemura1992}. The new La-Bi2201
data fits in a linear relation between relative intensity of SCP and
the optimal $T_c$ (Fig. 4f). Moreover, for the more stoichiometric
Bi$_{2}$Sr$_{2}$Ca$_{0.92}$Y$_{0.08}$Cu$_{2}$O$_{8+\delta}$ (OP96),
whose $T_c$ is enhanced to 96K from the usual optimal $T_c$ of 90K
\cite{Fujita2005,Eisaki2004}, its SCP intensity is enhanced, while
its gap is similar to that of OP90. Meanwhile, OP96 and Bi2223 have
similar SCP intensities, but Bi2223 has a larger gap. These
illustrate that improving either the gap size or phase coherence
would enhance $T_c$ \cite{Emery1995,DL2002}, and the low $T_c$ of
Bi2201 system is probably due to both its small gap size and low
superfluid density\cite{Eisaki2004,Uemura1992}.

To summarize, we have discovered the SCP in the antinodal region of
the single layer La-Bi2201.  In particular, we show that its 19\,meV
peak-dip separation intriguingly correlates with the energy scale of
spin fluctuations. Our data provides a critical piece to the global
picture, which would help to eventually resolve controversial issues
and uncover the ``glue'' of high-$T_c$ superconductivity.

We gratefully acknowledge the helpful discussions with Dr. W. S.
Lee, D. H. Lu, Profs. Z.-X. Shen, D. H. Lee, T. P. Devereaux and C.
Y. Kim. This work was supported by NSFC, MOST (973 project
No.2006CB921300 and No.2006CB601002), and STCSM of China. SSRL is
operated by the DOE Office of Basic Energy Science Divisions of
Chemical Sciences and Material Sciences.

\end{document}